# A New Security Mechanism for Vehicular Communication Networks

Ghassan Samara
Department of Computer Science, Faculty of Science and Information Technology, Zarqa University
Zarqa, Jordan.
gsamarah@yahoo.com

Wafaa A.H. Ali Alsalihy
School of Computer Science
Universiti Sains Malaysia
11800 Penang, Malaysia
Wafaa@cs.usm.my

*Abstract—Vehicular communication networks is a promising and emerging technology to facilitat road safety, Safety of life, traffic management, and infotainment dissemination for drivers and passengers. One of the ultimate goals in the design of such networking is to resist various malicious abuses and security attacks. In this research new security mechanisms are proposed to achieve secure certificate revocation, which is considered among the most challenging design objective in vehicular ad hoc networks.*

## I. INTRODUCTION

Traffic congestion is the most annoying thing that any driver in the world dreaming of avoiding it, a lot of traveling vehicles may cause problems, or facing problems that must be reported to other vehicles to avoid traffic overcrowding, furthermore, there are a lot of vehicles may send incorrect information, or a bogus data, and this could make the situation even worse.

Recent research initiatives supported by governments and car manufacturers seek to enhance the safety and efficiency of transportation systems. And one of the major topics to search is "Certificate Revocation".

Certificate revocation is a method to revoke some or all the certificates that the problematic vehicle has, this will enable other vehicles to avoid any information from those vehicles, which cause problems.

Current studies suggest that the Road Side Unit (RSU) is responsible for tracking the misbehavior of vehicles and for certificate revocation by broadcasting Certificate Revocation List (CRL). RSU also responsible for the certificate management, communication with Certificate Authority (CA), warning messages broadcasting, communicating with other RSUs. Figure 1 shows a type of RSU, where Mercedes Benz suggests that it is a small unit will be hanged on the street columns, every 1 KM [1] according to DSRC 5.9 GHZ range.

In vehicular ad hoc networks most of road vehicles will receive messages or broadcast sequence of messages, and they don't need to consider all of these Messages, because not all vehicles have a good intention and some of them have an Evil-minded.

Current technology suffers from high overhead on RSU, as RSU tacking responsibility for the whole Vehicular Network (VN) Communication. Furthermore, distributing CRL causes control channel consumption, as CRL need to be transmitted every 0.3 second [2]. Search in CRL for each message received causes a processing overhead for finding a single Certificate, where VN communication involves a kind of periodic message being sent and received 10 times per second.

This research proposes mechanisms that examine the certificates for the received messages, the certificate indicates to accept the information from the current Vehicle or ignore it; furthermore, this research will implement a mechanism for revoking certificates and assigning ones, these mechanisms will lead better and faster adversary vehicle recognizing.

## II. BACKGROUND RESEARCH

Existing works on vehicular network security [3], [4], [5], and [6] propose the usage of a PKI and digital signatures but do not provide any mechanisms for certificate revocation, even though it is a required component of any PKI-based solution.

In [7] Raya presented the problem of certificate revocation and its importance, the research discussed the current methods of revocation and its weaknesses, and proposed a new protocols for certificate revocation including : Certificate Revocation List (CRL), Revocation using Compressed Certificate Revocation Lists ($RC^2RL$), Revocation of the Tamper Proof Device (RTPD) and Distributed Revocation Protocol (DRP) stating the differences among them. Authors made a simulation on the DRP protocol concluding that the DRP protocol is the most convenient one which used the Bloom filter, the simulation tested a variety of environment like: Freeway, City and Mixing Freeway with City.

In [8] Samara divided the network to small adjacent clusters and replaced the CRL with local CRL exchanged interactively among vehicles, RSUs and CAs. The size of local CRL is small as it contains the certificates for the vehicles inside the cluster only.

In [9] Laberteaux proposed to distribute the CRL initiated by CA frequently. CRL contains only the IDs of misbehaving vehicles to reduce its size. The distribution of the received CRL from CA is made from RSU to all vehicles in its region, the problem of this method is that, not all the vehicles will receive the CRL (Ex: a vehicle in the Rural areas), to solve this problem the use of Car to Car (C2C) is introduced, using small number of RSU's, transmitting the CRL to the vehicles.

In [2] the eviction of problematic vehicles is introduced, furthermore, some revocation protocols like: Revocation of Trusted Component (RTC) and Leave Protocol are proposed.

In [10] some certificate revocation protocols were introduced in the traditional PKI architecture. It is concluded that the most commonly adopted certificate revocation scheme is through CRL, using central repositories prepared in CAs. Based on such centralized architecture, alternative solutions to CRL could be used for certificate revocation system like certificate revocation tree (CRT), the Online Certificate Status Protocol (OCSP), and other methods where the common requirement for these schemes is high availability of the centralized CAs, as frequent data transmission with On Board Unit (OBUs) to obtain timely revocation information may cause significant overhead.

### III. PROPOSED SOLUTION

In order to improve the security of certificate revocation problem, new protocols for message checking and certificate revocation will be proposed, the amount of invalid messages from bad attitude vehicles must minimized.

Message Checking:

In this approach any vehicle receives a message from any other vehicle takes the message and check for the sender certificate validity, if the sender has a Valid Certificate (VC), the receiver will consider the message, in contrary, if the sender has an Invalid Certificate (IC) the receiver will ignore the message, furthermore, if the sender doesn't have a certificate at all, the receiver will report to the RSU about the sender and check the message if it is correct or not, if the information received was correct RSU will give a VC for the sender, else RSU will give IC for it, and register the vehicle's identity into the CRL. See figure 2 and 3 for the VC and IC structure, and figure 4 for message checking.

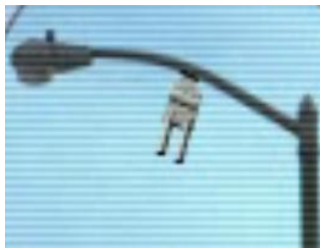

Figure1. Road side unit

| | |
|---|---|
| | Serial Number |
| | Issuer Name |
| Certificate Life Time → | Not Before |
| → | Not After |
| | Anonyms Name –if Any |
| | Public Key |

Figure 2. Valid certificate

IC will have the following fields:

| | |
|---|---|
| | Serial Number |
| | Issuer Name |
| | Reason of Revocation |
| Certificate Life Time → | Revocation Date |
| → | Review Date |
| | Anonyms Name –if Any |

Figure 3. Invalid certificate

These Certificates will not reserve much Memory like CRL does.

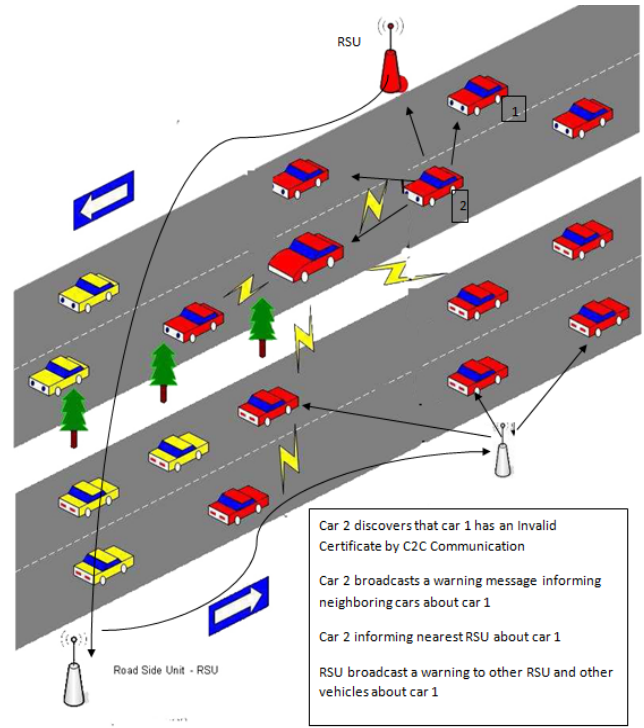

Figure 4. Message checking procedure

Certificate Revocation:

Certificate revocation is done when any misbehaving vehicle having VC is discovered, where RSU replaces the old VC with new IC, to indicate that this vehicle has to be avoided and this happens when more than one vehicle reporting to RSU that a certain vehicle has a VC and broadcasting wrong data. See figure 5, this report must be given to RSU each time that any receiver receives information from sender and finds that this information is wrong.

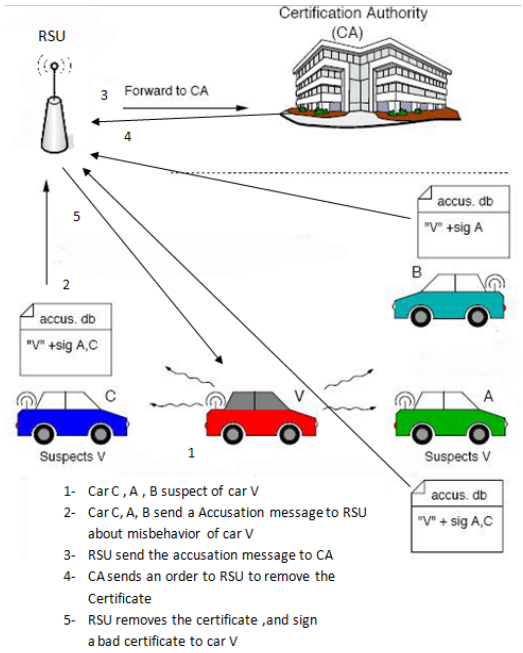

Figure 5. Certificate revocation procedure

The revocation will be as follows, a sender sen sends a message to receiver rec; this message may be from untrusted vehicle, so receiver sends Message to RSU to acquire Session Key (SKA), RSU replay message Containing SK Reply (SKR), this message contains the SK assigned to the current connection, this key is used to prevent attackers from fabrication of messages between the two vehicles.

Receiver sends a message to check validity, this message called "Validity Message", the message job is to indicate if the sender vehicle has a VC or not. Afterwards, RSU reports to the rec that the sender has a VC, so receiver can consider the information from the sender with no fear.

In some situations, receiver receives several massages, where all massages agree on a same result and same data, but a specific sender sends deferent data, this data will be considered as wrong data, if this data belongs to the same category.

Every message will be classified depending on its category:

TABLE I.   MESSAGE CLASSIFICATION AND CODING

| Code | Priority | Application |
|---|---|---|
| 001 | Safety of Life | Intersection Collision Warning/Avoidance |
| 002 | Safety of Life | Cooperative Collision Warning |
| 003 | Safety | Work Zone Warning |
| 004 | Safety | Transit Vehicle Signal Priority |
| 005 | Non-Safety | Toll Collection |
| 006 | Non-Safety | Service Announcement |
| 007 | Non-Safety | Movie Download( 2 hours of MPEG 1) |

Every category has a code, if the message received has the same code of the other messages, and has a deferent data, then this message is considered as a bogus message. In this case rec sends an Abuse Report (AR) for RSU, the Abuse AR (sen id, Message Code, Time of Receive), this report will be forwarded to CA, if RSU receives the same AR from other vehicles located in the same area, the number of abuse Report messages depends on the vehicles density on the road, see figure 6.

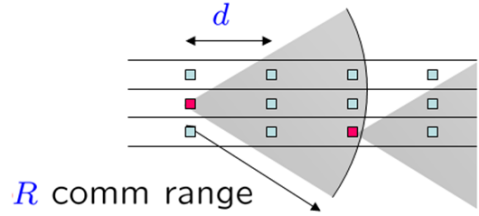

$l = $ # of lanes
$N = l \times R/d,$ # vehicles in range

Figure 6. Calculation of the Number of Vehicles in the Range [11].

If the number of vehicles that making accusation for a specific vehicle is near the half of the current vehicles, RSU will make a Revocation Request (RR) to revoke the VC from the sender vehicle.

Some vehicles don't produce an AR because they didn't receive any data from the sender vehicle (maybe they weren't in the area wile broadcasting), or they have a problem in their devices, or they have an IC, so RSU will not consider their messages.

CA makes a revocation order to RSU after confirming the RR and updates the CRL and then RSU revokes the VC from the sender vehicle, and assigns IC for it, to indicate to other vehicles in the future, that this vehicle broadcasts wrong data, "don't trust it".

Figure 5 shows certificate revocation steps.

The message sequence is as follows

1- sen ⟶ rec:ENCPK[Message + Sig]

2- rec ⟶ RSU:ENCPK [SKA + Sig]

3- RSU ⟶ rec: ENCPK [SKR + T + Sig]

4- rec ⟶ sen: ENCSK [VM + T + Sig]

5- sen ⟶ RSU:ENCPK [SKA + Sig]

6- RSU ⟶ sen: ENCPK [SKR + T + Sig]

7- sen ⟶ rec: ENCSK [VC + T + Sig]

8- rec ⟶ RSU: ENCPK [AbR + T + Sig]

9- RSU ⟶ CA: ENCPK [SN + RevReq + RevRea+ T + Sig + senid + MesCod]

10- CA ⟶ RSU: ENCPK [SN + RevOr + UpCRL + T + Sig + senid + RevRea]

11- RSU ⟶ sen: ENCPK [RevVC + AssBC + T + Sig + RevRea]

Message 1: sen (sender) sends a message to the rec (receiver), this message along with digital signature of sen, and this message is encrypted with the Primary Key (PK) of rec.

Any attacker can make a fabricated message telling rec that this message originated from sen, to prevent this signature from being used.

Message 2: rec sends a request to RSU encrypted with the PK of RSU, acquiring a SK for securing connection.

Message 3: replay for Message 2, contains the SK and the time for sending the replay, the importance of the time is to prevent replay attack, where an attacker can send this message more than once, with the same session key, and same signature, so he can forge the whole connection.

Message 4: rec sends validity message to check if the vehicle has to be avoided or not, this message encrypted with the shared SK obtained from RSU.

Message 7: sen sends a message to rec containing the VC, to report for rec that this vehicle must be trusted, and the time of sending, in here, to avoid reply attack, which happens when an attacker keeps the message with him, and sends it after a period, may be at that time, the senders certificate been revoked by RSU, so the sen must be avoided, but the attacker force the rec vehicle to trust it. After receiving the information, rec checks if the message has a deferent or same data for the same category of other messages received.

Message 8: if the message is deferent, then, wrong data is received, rec sends an Abuse Report for RSU, contains sen id to know which vehicle made the problem, Message Code to know the category of the message, Time of Receive to know when the message received, and the message also includes the Time to avoid replay attack and Signature to avoid fabrication; the message is encrypted with PK of RSU.

In this situation replay attack will happen, if an attacker copied this message, and sends it frequently to RSU in several times to make sure that the number of accusation reached a level, that the certificate must be revoked.

After examining the number of vehicles that accused sen for sending an Invalid message, if the number is reasonable, RSU sends Message 9.

Message 9: RSU sends RR for CA, containing Serial Number and Time to avoid replay attack and Signature to avoid fabrication, Revocation Reason to state what is the reason for revocation, and sen id to know which vehicle is the problematic one and message code to know what is the message category; the message is encrypted with PK of CA.

Replay attack in this situation happens when an attacker wants to transmit the same message for CA claiming that this message is from RSU, after some time CA will not have the ability to respond, causing for DoS attack, so RSU must use Time and Serial number for this message, because CA has a lot of work to do and sending a lot of these kind of messages will cause a problem.

Message 10: CA makes a Revocation Order for RSU; this message contains SN to avoid DoS Attack, time to avoid replay attack, signature to avoid fabrication attack, Sender Id, Revocation Reason to state what is the reason for revocation.

After receiving this request CA will update CRL, adding the new vehicle that been captured to CRL and send it for RSU.

DoS attack can happen, when attacker keep sending the same message to RSU, claiming that the message originated from CA, CA messages have the highest priority to be processed by RSU, so RSU will receive a huge amount of messages from CA and process it, without having the time to communicate with other RSUs or other vehicles, to avoid it a serial number and signature is used.

Message 11: RSU makes the revocation, revoking VC, assigning IC, also this message contains the time to avoid replay attack, Signature to avoid fabrication attack, Revocation Reason to state what is the reason for revocation.

However, RSU will be responsible for renewing vehicle certificates, any vehicle has an expiring certificate will communicate with RSU to renew the certificate, then the RSU will check the CRL to see if this vehicle has an IC or not. If there is no problem for giving a new certificate for this vehicle, it will be given for a specific life time, when the period expires vehicle will issue a request for the CA for renewing the certificate. VC will have a special design different from the design of X.509 certificate [12] as shown in figure 2 and 3.

## IV. RESULTS AND DISCUSSION

CRL is the most common mechanism for certificate revocation in VANET, in [13] authors proposed that each vehicle must be stored with approximately 25000 certificates, and each certificate has 100 bytes, so, 2.5 mb is required to store the revocation data for one vehicle in CRL. By simple calculation the size of the CRL having 100 adversary vehicles might be 250 mb, and this is very large size to be broadcasted frequently for dense and high mobile network like VANET.

The proposed mechanism would replace the old one, where the adversary vehicles would be identified by VC and IC certificates, so the size of the CRL will be reduced 90 % and this makes fast and efficient distribution and adversary recognition in VANET.

## V. CONCLUSION

CRL is considered as the most common mechanism for adversary recognition in VANET which causes long delay, processing overhead and channel jamming. In this research new security mechanisms were proposed to achieve secure certificate revocation, which is considered among the most challenging design objective in vehicular ad hoc networks.

The proposed mechanisms help vehicles to easily identify the adversary vehicle and make the certificate revocation for better certificate management.

## VI. ACKNOWLEDGMENT

This work is part of Universiti Sains Malaysia short term grant No. 304/PKMP/6363/1090.